**Safety First: Psychological Safety as the Key to AI Transformation**

Aaron Reich[1], Diana Wolfe[2], Matt Price[3], Alice Choe[4], Fergus Kidd[5], Hannah Wagner[6]

[1] Chief Technology and Innovation Officer, Avanade

[2] PhD in Industrial Organizational Psychology; Vice President, Head of AI Research and Strategy, Kyndryl

[3] PhD in Media Psychology & Behavioral Neuropsychology; AI Governance, Responsible AI and Emerging Technology Platform Lead, Avanade

[4] PhD Candidate, Organizational Behavior and Human Resource Management, Rotman School of Management, University of Toronto

[5] Chief Technology Office, FieldPal.ai

[6] PhD Candidate, Industrial Organizational Psychology, Seattle Pacific University, 3307 3rd Ave. W, Seattle, WA

Correspondence concerning this article should be addressed to: aaron.reich@avanade.com

**Abstract**

Organizations continue to invest in artificial intelligence (AI), yet many struggle to ensure that employees adopt and engage with these tools. Drawing on research highlighting the interpersonal and learning demands of technology use, this study examines whether psychological safety (i.e., employees' belief that they can ask questions, experiment, and make mistakes without fear of negative consequences) is associated with AI adoption and usage in the workplace. Using survey data from 2,257 employees in a global consulting firm, we test whether psychological safety is associated with (a) adoption, (b) usage frequency, and (c) usage duration, and whether these relationships vary by organizational level, professional experience, or geographic region. Logistic and linear regression analyses show that psychological safety reliably predicts whether employees adopt AI tools but does not predict how often or how long they use AI once adoption has occurred. Moreover, the relationship between psychological safety and AI adoption is consistent across experience levels, role levels, and regions, and no moderation effects emerge. These findings suggest that psychological safety functions as a key antecedent of initial AI engagement but not of subsequent usage intensity. The study underscores the need to distinguish between adoption and sustained use and highlights opportunities for targeted organizational interventions in early-stage AI implementation.

**Safety First: Psychological Safety as the Key to AI Transformation**

Organizations across industries continue to invest heavily in artificial intelligence (AI), yet many struggle with the human factors that determine whether employees will actually adopt and use these tools (Kemp, 2025). While traditional implementation frameworks emphasize technological readiness and task alignment, emerging scholarship highlights the importance of psychological conditions, particularly the degree to which employees feel safe experimenting with unfamiliar tools, as a determinant of AI engagement (Kim et al., 2025).

Organizational research has long shown that adopting new technologies requires individuals to navigate uncertainty, surface skill gaps, and learn through trial-and-error. Classic work on technological change demonstrates that early stages of adoption often involve iterative experimentation and reconfiguration of work practices (Lewis et al., 2003; Tyre & Orlikowski, 1994). These learning processes inherently expose employees to the possibility of mistakes and social evaluation, suggesting that psychological safety may be especially important in AI implementation contexts.

Psychological safety—an individual's belief that they can ask questions, seek help, and admit mistakes without fear of negative consequences (Edmondson, 1999)—has been identified as a core facilitator of learning behavior, idea sharing, and experimentation (Frazier et al., 2017). Because AI tools often produce unpredictable outputs and demand new competencies, employees who lack psychological safety may avoid engaging with these systems, whereas those who feel safe may be more willing to explore, troubleshoot, and learn. Industry reports similarly emphasize that psychological safety reduces fear of failure during digital transformation and encourages employees to experiment with emerging technologies (Deloitte, 2021).

The present study tests this possibility within a global consulting organization. Using survey data from 2,257 employees, we examine whether individual perceptions of psychological safety predict (a) AI adoption, (b) usage frequency, and (c) usage duration. We also investigate whether these relationships vary across role level, professional experience, and geographic region. This approach enables a deeper understanding of the psychological conditions that support AI engagement in complex, knowledge-intensive environments.

**Theoretical Background**

Psychological safety is a well-established predictor of behaviors central to learning and adaptation at work. Edmondson's (1999) foundational research demonstrated that individuals who feel psychologically safe are more likely to speak up, ask questions, report errors, and seek feedback. Importantly, psychological safety varies within teams: even when shared norms exist, individuals differ in the extent to which they personally feel able to take interpersonal risks. A meta-analysis of 117 studies further confirms that psychological safety predicts information sharing, learning behavior, and creativity (Frazier et al., 2017).

These mechanisms are directly applicable to AI engagement. Unlike many prior technologies, AI tools often produce uncertain, ambiguous, or imperfect outputs. Using them effectively requires employees to test boundaries, troubleshoot errors, and—crucially—reveal gaps in their own technological fluency. Individuals who fear negative evaluation may avoid such exposure, consistent with research showing that perceived interpersonal risk reduces experimentation and willingness to surface errors (Edmondson, 1999). In contrast, psychologically safe environments allow employees to "fail forward" and learn from mistakes, a dynamic emphasized in both academic studies and industry analyses of digital transformation (Deloitte, 2021).

Recent empirical work supports the relevance of psychological safety in AI contexts. Kim et al. (2025) find that psychological safety buffers employees from anxiety and withdrawal during rapid AI implementation. Shen et al. (2025) show that teachers with higher psychological safety and trust are more likely to continue using AI tools over time. These findings indicate that psychological safety not only facilitates initial adoption but also supports ongoing engagement with AI systems.

Conceptualizing psychological safety at the individual level aligns with these processes: adopting AI requires personal judgment, willingness to reveal uncertainty, and tolerance for ambiguous outcomes. While organizational practices shape the broader climate, the decision to try an AI tool is ultimately an individual-level psychological act. This distinction suggests that psychological safety should be most influential at the point of initial adoption, when uncertainty and interpersonal risk are highest, whereas usage frequency and duration may depend more on habit formation, workflow integration, and perceived usefulness.

Together, this literature motivates the present study's examination of psychological safety as a predictor of AI engagement and its potential moderators across employee experience, organizational level, and geographic region.

**Research Questions and Hypotheses**

The present study investigates whether employees' perceptions of psychological safety predict meaningful differences in AI engagement within a large, multinational consulting organization. Drawing on research emphasizing the importance of a supportive interpersonal climate in facilitating risk-taking, experimentation, and learning (Edmondson, 1999; Frazier et al., 2017), this study examines psychological safety as a potential antecedent to three core AI behaviors: (a) adoption, (b) usage frequency, and (c) usage duration. Unlike research that focuses solely on technology characteristics or individual attitudes, this study centers on a construct that has been repeatedly linked to exploratory behavior and innovation: psychological safety. The overarching goal is to determine whether psychological safety provides the conditions under which professionals feel more willing to try AI tools in the first place and whether they continue to shape usage patterns among adopters. In addition, the study explores whether the psychological safety–AI relationship depends on employees' years of experience, organizational level, or geographic region.

### *Psychological Safety and AI Engagement*

Psychological safety supports interpersonal risk-taking, learning behavior, and experimentation, all of which are central to early-stage technology engagement (Edmondson & Lei, 2014). When individuals feel safe to make mistakes or ask questions, they are more likely to explore new tools and share emerging practices. AI tools often involve uncertain outcomes, iterative use, and the potential for errors, suggesting that psychological safety may play a central role in enabling adoption. However, once employees have incorporated AI tools into their workflow, usage frequency and duration may be driven more by role demands, task relevance, and workflow integration than by psychological safety alone. Given this distinction between adoption and usage intensity, the following hypotheses are proposed:

**H1a:** Higher perceived psychological safety will be associated with greater likelihood of AI adoption.

**H1b:** Among employees who have adopted AI tools, higher perceived psychological safety will predict more frequent AI use.

**H1c:** Among employees who have adopted AI tools, higher perceived psychological safety will predict longer duration of AI use (sustained use over time).

*Years of Experience as a Moderator*

Employees with different levels of professional experience may draw on distinct learning histories, norms, and coping strategies when interpreting psychologically safe or unsafe environments. Experience influences how individuals make sense of workplace cues and how they navigate new or uncertain situations, including those involving technology use (Tesluk & Jacobs, 1998; Louis, 1980). Prior technology acceptance research suggests that experience shapes responses to determinants such as social influence, perceived ease of use, and perceived behavioral control (Venkatesh et al., 2003). Professionals with extensive experience may approach new tools through well-developed routines and expectations, whereas those earlier in their careers may rely more heavily on ongoing interpersonal feedback and contextual signals. Given these differing frames of reference, professional experience may shape how psychological safety relates to AI engagement, although the nature of this relationship remains an open empirical question.

> **H2a:** Professional experience will moderate the association between psychological safety and AI adoption.
>
> **H2b:** Professional experience will moderate the association between psychological safety and AI usage frequency.
>
> **H2c:** Professional experience will moderate the association between psychological safety and AI usage duration.

*Organizational Level as a Moderator*

Organizational level reflects differences in authority, autonomy, visibility, and performance expectations (Katz & Kahn, 1978; Magee & Galinsky, 2008). Senior employees often perceive greater latitude to experiment with new tools regardless of interpersonal risk because higher-status roles afford increased psychological and structural safety (Fast et al., 2012). Junior employees, in contrast, are typically more sensitive to interpersonal risk and social evaluation due to lower status, greater evaluative scrutiny, and fewer control resources (Kim et al., 2019; Anderson et al., 2012). Prior research suggests that hierarchical position can shape technology attitudes and uptake opportunities, influencing technostress exposure, perceived control, and access to support (Tarafdar et al., 2010). Yet it remains unclear whether organizational level changes how psychological safety relates to AI behavior. Therefore, the following hypotheses assess whether psychological safety operates consistently across levels or whether its influence varies depending on organizational position:

> **H3a:** Organizational level will moderate the association between psychological safety and AI adoption.
>
> **H3b**: Organizational level will moderate the association between psychological safety and AI usage frequency.
>
> **H3c**: Organizational level will moderate the association between psychological safety and AI usage duration.

***Geographic Region as a Moderator***

Geographic region encompasses cultural, institutional, and organizational contexts that shape how employees interpret interpersonal cues and risk-taking norms (House et al., 2004; Hofstede, 2001). Cross-cultural research on technology adoption indicates that cultural values such as uncertainty avoidance, power distance, and institutional trust influence individuals' willingness to experiment with new technologies and rely on digital systems (Im et al., 2011; Srite & Karahanna, 2006; Straub et al., 1997). These cultural differences also affect perceptions of error tolerance, psychological safety, and acceptable forms of workplace communication (Gelfand et al., 2007; Brockner et al., 2001). As a result, the degree to which psychological safety supports AI experimentation may vary across cultural contexts. Although adoption rates for digital tools can appear similar across regions within multinational firms, the *meaning* and *interpersonal consequences* of taking risks—such as asking questions, admitting uncertainty, or experimenting with imperfect systems—can differ substantially (Taras et al., 2016). Given this cross-cultural variability, the following hypotheses treat regional differences as exploratory:

**H4a:** Geographic region will moderate the association between psychological safety and AI adoption.

**H4b:** Geographic region will moderate the association between psychological safety and AI usage frequency.

**H4c:** Geographic region will moderate the association between psychological safety and AI usage duration.

Together, these hypotheses evaluate the degree to which psychological safety serves as a key antecedent to AI adoption and whether this relationship generalizes across professional experience levels, organizational positions, and global regions.

**Methods**

*Participants and Procedure*

Data were collected from 2,257 professionals across a single multinational consulting organization through Microsoft Forms, an online survey platform. Email invitations were distributed to all eligible participants, with reminders sent to increase response rates. The consulting context provides an appropriate setting for examining AI acceptance, as knowledge workers in this industry represent early adopters of cognitive technologies while facing direct implications of AI capabilities for their professional roles. The multinational scope enables examination of both organizational hierarchy and geographic effects within a consistent industry context.

Participants represented diverse organizational levels, geographic regions, and experience levels. Organizational levels included analyst and senior analyst, consultant and senior consultant, manager and group manager, director and senior director, and executive positions. Geographic regions encompass Europe, Growth Markets, North America, and the Global Delivery Network (GDN). Experience levels were categorized into: less than one year, 1-3 years, 3-5 years, 5-10 years, and 10 or more years. The survey was administered in English and took approximately 25 minutes to complete. Of the total sample, 1,256 participants (55.7%) reported current use of AI tools in their work, while 1,001 (44.3%) reported no current AI tool usage.

*Measures*

**AI Adoption and Usage Intensity.** AI adoption was assessed with a single dichotomous item asking participants whether they used any AI tools in their regular workflow ("Do you use AI tools in your workflow?"; response options: *yes* or *no*). Among AI users, two additional items captured usage intensity. Frequency of use was measured on a six-point ordinal scale ranging from *0 = on rare occasions* to *5 = multiple times a day.* Duration of use captured the length of time respondents had incorporated AI into their workflow, rated on a six-point scale from *0 = less than one month* to *5 = over two years.* These items provided a basic behavioral profile of participants' AI adoption patterns.

**Psychological Safety.** Psychological Safety. Psychological safety was assessed using three dimensions based on Edmondson's (1999) framework. Individual Psychological Safety (4 items) measured employees' comfort with interpersonal risk-taking; a sample item was "I feel safe offering new ideas, even if they aren't fully-formed ideas." Team Respect (3 items) assessed perceptions of acceptance and value for diverse contributions; a sample item was "Members of my work team could easily describe the value of others' contributions." Team Learning (4 items) captured the team's orientation toward learning from mistakes and continuous improvement; a sample item was "Members of my work team raise concerns they have about team plans or decisions." All items were rated on seven-point Likert scales anchored at 1 = strongly disagree and 7 = strongly agree. A composite psychological safety score was computed as the mean across all 11 items. Full scale items are presented in the Appendix.

**Demographic Variables.** Participants reported several demographic and professional characteristics. Years of professional experience was measured categorically, with five levels ranging from less than one year to more than ten years. Organizational level included four categories after excluding executives due to small sample size. Geographic region reflected four major business regions within the organization: Europe, Growth Markets, North America, and the Global Delivery Network.

## Results

### Descriptives

As shown in Table 1, AI adoption rates were relatively consistent across experience groups, ranging from 53.5% among those with 10+ years of experience to 62.0% among those with less than one year. Table 2 displays adoption rates by organizational level, with usage ranging from 50.0% among Consultants/Sr Consultants to 70.4% among Directors/Sr Directors. Table 3 summarizes adoption by region, with rates spanning from 54.9% in Europe to 57.5% in North America.

These descriptive statistics indicate that AI tool adoption is broadly distributed across demographic groups, with no single group showing markedly higher or lower usage.[1] This pattern provides a robust foundation for examining whether psychological safety and its potential moderators predict meaningful differences in AI engagement across the workforce.

**Table 1**

*AI Usage Patterns by Years of Experience*

| Years of Experience | Total N | AI Non-Users (n) | AI Users (n) | Usage Rate (%) |
|---|---|---|---|---|
| Less than one year | 142 | 54 | 88 | 62.0% |
| 1-3 years | 365 | 169 | 196 | 53.7% |
| 3-5 years | 310 | 122 | 188 | 60.6% |
| 5-10 years | 436 | 189 | 247 | 56.7% |
| 10+ years | 997 | 464 | 533 | 53.5% |

**Table 2**

*AI Usage Patterns by Organizational Level*

| Organizational Level | Total N | AI Non-Users (n) | AI Users (n) | Usage Rate (%) |
|---|---|---|---|---|
| Analyst/Sr Analyst | 620 | 264 | 356 | 57.4% |
| Consultant/Sr Consultant | 836 | 418 | 418 | 50.0% |
| Manager/Group Manager | 643 | 271 | 372 | 57.9% |
| Director/Sr Director | 142 | 42 | 100 | 70.4% |

---

[1] We do not present detailed analyses of AI usage frequency and duration by organizational level, years of experience, or region in this paper. These analyses, conducted on the same sample, are reported in our recent publication, "Revisiting UTAUT for the Age of AI: Understanding Employees' AI Adoption and Usage Patterns Through an Extended UTAUT Framework", now available on arXiv (https://doi.org/10.48550/arXiv.2510.15142). In that work, we found that demographic differences in AI usage frequency and duration were minimal and generally not statistically significant. Readers interested in a comprehensive breakdown of usage patterns by demographic group are encouraged to consult that paper for full results and interpretation.

*Note.* Executive level excluded due to small sample size (n < 10).

**Table 3**

*AI Usage Patterns by Region*

| Region | Total N | AI Non-Users (n) | AI Users (n) | Usage Rate (%) |
|---|---|---|---|---|
| Europe | 1016 | 458 | 558 | 54.9% |
| GDN | 37 | 16 | 21 | 56.8% |
| Growth Markets | 629 | 281 | 348 | 55.3% |
| North America | 562 | 239 | 323 | 57.5% |

Nonparametric tests revealed no statistically significant differences in psychological safety across demographic groups (see Table 4 for descriptives). Kruskal-Wallis tests for Years of Experience, Current Level, and Region yielded p-values well above the Bonferroni-adjusted threshold ($\alpha = 0.017$), and all effect sizes ($\varepsilon^2$) were negligible (< 0.01). These findings indicate that perceptions of psychological safety, whether at the individual level or in terms of team respect and learning, are broadly consistent across tenure, organizational level, and geographic region.

**Table 4**

*Psychological Safety by Demographic Variables*

| | Category | n | M | SD |
|---|---|---|---|---|
| Overall Sample | | 2,209 | 3.91 | 0.61 |
| Role Level | Analyst/Sr Analyst | 609 | 3.91 | 0.63 |
| | Consultant/Sr Consultant | 821 | 3.90 | 0.59 |
| | Director/Sr Director | 141 | 4.00 | 0.55 |
| | Manager/Group Manager | 638 | 3.89 | 0.63 |
| Years of Experience | 1–3 years | 359 | 3.94 | 0.61 |
| | 10+ years | 982 | 3.91 | 0.61 |
| | 3–5 years | 303 | 3.86 | 0.64 |
| | 5–10 years | 426 | 3.91 | 0.61 |
| | Less than one year | 139 | 3.87 | 0.53 |
| Region | Europe | 997 | 3.93 | 0.60 |
| | GDN | 35 | 3.88 | 0.70 |
| | Growth Markets | 623 | 3.87 | 0.62 |
| | North America | 554 | 3.90 | 0.62 |

*Note.* M = mean on 5-point scale (1 = strongly disagree, 5 = strongly agree); SD = standard deviation. Scores reflect mean composite ratings. Higher scores indicate *greater* felt psychological safety.

### Reliability and Normality Assessment

Prior to hypothesis testing, we assessed the psychometric properties of our key measures. Cronbach's alpha analyses revealed that psychological safety scales demonstrated acceptable to good internal consistency: Individual Psychological Safety ($\alpha$ = .79), Team Respect ($\alpha$ = .85), and Team Learning ($\alpha$ = .87).

Normality testing using Shapiro-Wilk tests indicated significant departures from normal distributions for all psychological safety dimensions (all ps < .001), with test statistics ranging from W = 0.915 to W = 0.951. These violations of normality assumptions are common in Likert-scale survey data due to response patterns such as ceiling effects and social desirability bias. Given these distributional characteristics, we employed robust analytical approaches throughout our analyses: logistic regression for binary adoption outcomes (which does not assume normality of predictors) and ordinary least squares regression with robust standard errors for continuous outcomes. This analytical strategy ensures that our statistical inferences remain valid despite the non-normal distributions observed in our psychological safety measures.

*Main effects (H1)*

Psychological safety significantly predicted whether employees adopted AI tools (see Table 5). Logistic regression showed that higher psychological safety increased the likelihood of adoption, β = 0.26, SE = 0.07, p < .001, 95% CI [0.12, 0.40]. The corresponding odds ratio indicated that each one-unit increase in psychological safety was associated with a 29.6% increase in the odds of AI adoption, OR = 1.30, 95% CI [1.13, 1.49]. The corresponding odds ratio indicated that each one-unit increase in psychological safety was associated with a 29.6% increase in the odds of AI adoption, OR = 1.30, 95% CI [1.13, 1.49]. To contextualize this effect, meta-analyses of technology adoption find that core UTAUT predictors such as performance expectancy typically yield standardized coefficients of β = .30 or higher, while secondary predictors like social influence and facilitating conditions range from β = .10 to .20 (Venkatesh et al., 2003; Dwivedi et al., 2020). The effect of psychological safety on AI adoption falls within this latter range, suggesting it operates as a meaningful but secondary antecedent—comparable in magnitude to social and contextual factors rather than to perceived usefulness or performance gains.

However, among employees who had adopted AI tools, psychological safety did not significantly predict how frequently they used AI, β = 0.08, SE = 0.05, p = .13, 95% CI [-0.02, 0.18]. Psychological safety was also unrelated to the duration of AI use. The effect was near zero, β = 0.01, SE = 0.06, p = .86, 95% CI [-0.10, 0.12].

Together these findings suggest that perceived psychological safety influences the decision to try AI tools but does not meaningfully influence how extensively employees use AI once they have adopted it.

**Table 5**

*Main Effects of Psychological Safety on AI Adoption and Usage Outcomes*

| Outcome Variable | n | β (SE) | 95% CI | OR | p |
|---|---|---|---|---|---|
| Adoption (Logistic) | 2,228 | 0.259 (0.070) | [0.121, 0.397] | 1.296 | <.001 |
| Frequency (Linear)[a] | 1,232 | 0.078 (0.051) | [-0.023, 0.179] | *n/a* | *n/a* |
| Duration (Linear)[a] | 1,233 | 0.010 (0.057) | [-0.102, 0.122] | *n/a* | *n/a* |

Note. β = unstandardized regression coefficient; SE = standard error; CI = confidence interval; OR = odds ratio for logistic regression. [a]Analysis conducted among AI users only.

### Testing Moderation by Role Level (H2)

To test whether organizational role level moderates the relationship between psychological safety and AI adoption (H2a), we conducted a logistic regression analysis with robust standard errors (HC3). The model included psychological safety as the predictor, role level dummy variables (with Analyst/Sr Analyst as the reference category), and their interaction terms.

Role level did not significantly moderate the relationship between psychological safety and AI outcomes (see Table 6). Likelihood ratio tests for interaction terms were nonsignificant for AI adoption ($\chi^2(3) = 2.66$, p = .45), AI use frequency ($\Delta LL = 0.53$, p = .91), and AI use duration ($\Delta LL = 3.40$, p = .33). This indicates that the effect of psychological safety on AI-related behaviors is consistent across Analyst, Consultant, Manager, and Director roles.

For AI adoption (H2a), psychological safety remained a significant predictor ($\beta = 0.29$, $SE = 0.13$, $p = 0.025$, OR = 1.34, 95% CI [1.04, 1.72]), meaning a one-unit increase in psychological safety increases the odds of AI adoption by 33.6% for the reference group (Analyst/Sr Analyst). However, for AI use frequency (H2b) and AI use duration (H2c) among existing AI users, psychological safety was not significant (p = .20 and p = .82, respectively).

Overall, psychological safety matters for whether employees start using AI, but not for how often or how long they use it. Role level does not alter these relationships, suggesting that interventions to enhance psychological safety should focus on encouraging initial adoption rather than influencing usage intensity.

**Table 6**

*Moderator Tested: Role Level*

| Hypothesis | Analysis | N | ΔLL | df | p |
|---|---|---|---|---|---|
| H2a: Role × PS → AI Adoption | Logistic Regression | 2,220 | 2.66 | 3 | .45 |
| H2b: Role × PS → AI Frequency | OLS Regression | 1,228 | 0.53 | 3 | .91 |
| H2c: Role × PS → AI Duration | OLS Regression | 1,229 | 3.40 | 3 | .33 |

Note. N = sample size; ΔLL = change in log-likelihood; df = degrees of freedom; p = p-value.

### Testing Moderation by Experience (H3)

Experience level did not significantly moderate the relationship between psychological safety and AI adoption (see Table 7). The interaction term was nonsignificant ($\beta = 0.01$, $SE = 0.06$, $p = 0.83$, 95% CI [-0.10, 0.12]), indicating no moderation effect. Psychological safety remained a significant predictor of adoption ($\beta = 0.26$, $SE = 0.07$, $p < 0.00$), while experience level showed no significant main effect ($\beta = -0.10$, $SE = 0.22$, $p = 0.65$).

Neither main effect was statistically significant for AI frequency: psychological safety ($\beta = 0.08$, SE = 0.05, p = 0.15, 95% CI [-0.03, 0.18]) and experience level ($\beta = -0.02$, SE = 0.17, p = 0.92, 95% CI [-0.35, 0.31]). The interaction between psychological safety and experience level was not statistically significant ($\beta = 0.01$, $SE = 0.04$, $p = 0.90$, 95% CI [-0.08, 0.09]). The F-test for the interaction term confirmed no significant moderation effect ($p$ = 0.89), indicating that experience level does not moderate the relationship between psychological safety and AI use frequency.

Similarly, psychological safety ($\beta$ = 0.01, *SE* = 0.06, $p$ = 0.91, 95% CI [-0.11, 0.12]) and experience level ($\beta$ = 0.04, *SE* = 0.18, $p$ = 0.82, 95% CI [-0.32, 0.40]) were not significant predictors of AI use duration, and the interaction term was nonsignificant ($\beta$ = 0.01, *SE* = 0.05, $p$ = 0.78, 95% CI [-0.08, 0.10]). Psychological safety influences adoption but not usage depth, regardless of experience level.

**Table 7**

*Moderator Tested: Years of Experience*

| Hypothesis | Analysis | N | Psych Safety × Experience ($\beta$) | p |
|---|---|---|---|---|
| H3a: Experience × PS → AI Adoption | Logistic Regression | 2,229 | .0116 | .83 |
| H3b: Experience × PS → AI Frequency | OLS Regression | 1,230 | 0.0057 | .89 |
| H3c: Experience × PS → AI Duration | OLS Regression | 1,231 | 0.0127 | .78 |

### Testing Moderation by Region (H4)

To examine whether geographic region moderates the relationship between psychological safety and AI adoption (H4a), we conducted a logistic regression analysis with HC3 robust standard errors. The analysis included 2,224 participants across four regions: Europe (45.1%, 54.5% adoption), Growth Markets (28.2%, 55.2% adoption), North America (25.0%, 57.3% adoption), and GDN (1.7%, 56.8% adoption).

Region did not significantly moderate the relationship between psychological safety and AI adoption. The overall interaction test was nonsignificant ($\Delta$LL = 5.78, p = .22). Among AI users, psychological safety did not predict usage frequency ($\beta = 0.03$, $p = .72$), and region interactions were also nonsignificant (overall F-test $p = .69$).

**Discussion**

The present study demonstrates a clear and consistent pattern: psychological safety is a significant predictor of whether employees adopt AI tools in the workplace. Across a large, multinational consulting sample, individuals reporting higher psychological safety were substantially more likely to engage with AI, even after accounting for differences in organizational level, professional experience, and geographic region. This effect was robust and remained significant regardless of demographic context, underscoring psychological safety as a foundational driver of early-stage AI engagement.

In contrast, psychological safety did not predict how frequently or how long employees used AI once they had adopted it. Usage frequency and duration were largely uniform across demographic groups, suggesting that the factors influencing initial adoption differ from those shaping ongoing engagement. This distinction supports the idea that adoption and usage represent distinct behavioral processes: while adoption depends heavily on willingness to experiment and take interpersonal risks, usage intensity may be governed by a distinct set of drivers. Stage models of technology acceptance suggest that once users cross the adoption threshold, behavior becomes increasingly shaped by perceived usefulness, task-technology fit, and habit formation (Venkatesh et al., 2003; Limayem et al., 2007).

At this point, employees are no longer deciding whether to try the tool; they are evaluating whether it makes their work faster, easier, or better. These instrumental considerations, along with the gradual routinization of AI into daily workflows, may supersede the interpersonal risk calculus that psychological safety addresses. Future research should examine how these post-adoption factors interact with initial safety perceptions to shape sustained AI engagement. Future research should examine how alternative factors shape sustained AI engagement.

Notably, our findings diverge from Shen et al. (2025), who found that psychological safety predicted teachers' continued use of AI tools in classroom settings. Several factors may account for this discrepancy. First, the populations differ substantially: teachers operate in highly visible, evaluative environments where ongoing AI use remains interpersonally risky (e.g., parents questioning AI-assisted grading), whereas consultants may face reduced scrutiny once AI is integrated into individual workflows. Second, Shen et al. measured continuance intention rather than actual usage behavior—and intention-behavior gaps are well-documented in technology acceptance research (Venkatesh et al., 2003). Third, their model included trust as a parallel predictor; trust may account for variance in continued use that psychological safety alone does not capture. Future research should examine boundary conditions for the psychological safety–usage relationship across occupational contexts and measurement approaches.

The present study demonstrates that psychological safety is a significant predictor of whether employees adopt AI tools in the workplace, with higher levels of psychological safety consistently associated with greater odds of adoption. This effect was robust across organizational level, professional experience, and geographic region, suggesting psychological safety is a foundational driver of early-stage AI engagement. In contrast, psychological safety did not predict how frequently or how long employees used AI once they had adopted it, indicating that adoption and usage represent distinct behavioral processes.

***Limitations***

Several limitations should be acknowledged. First, the cross-sectional design precludes causal inference; while psychological safety is associated with AI adoption, we cannot determine whether it directly causes increased adoption or whether other unmeasured factors contribute to this relationship. Second, all data were self-reported, which may introduce bias due to social desirability or recall error. Third, the study was conducted within a single multinational consulting organization, which may limit the generalizability of findings to other industries or organizational cultures. Finally, while we tested for moderation by role, experience, and region, other contextual factors (e.g., team climate, leadership style, or organizational readiness) may also shape technology engagement and were not assessed here. Finally, we did not control for individual differences in AI self-efficacy, prior technology experience, or perceived usefulness, all of which may independently influence adoption decisions.

### *Practical Implications*

Rather than prescribing specific interventions, these findings highlight the value of an experimental approach for organizations seeking to accelerate technology adoption and usage. By systematically testing whether factors like psychological safety matter (or do not matter) for different stages of technology engagement, organizations can more precisely identify where barriers exist. This evidence-based strategy enables leaders to target resources and interventions where they will have the greatest impact, whether that means fostering psychological safety to encourage initial adoption or addressing workflow integration and training to support sustained use. Ultimately, an experimental mindset helps organizations move beyond assumptions and focus on the real drivers of successful technology transformation.

**References**

Anderson, C., Kraus, M. W., Galinsky, A. D., & Keltner, D. (2012). The local-ladder effect: Social status and subjective well-being. Psychological Science, 23(7), 764–771. https://doi.org/10.1177/0956797611434537

Brill, J. (2025, September 22). Psychological safety drives AI adoption. Psychology Today. Retrieved from https://www.psychologytoday.com

Brockner, J., Ackerman, G., Greenberg, J., Gelfand, M. J., Francesco, A. M., Chen, Z. X., ... & Shapiro, D. (2001). Culture and procedural justice: The influence of power distance on reactions to voice. Journal of Experimental Social Psychology, 37(4), 300–315.

Deloitte. (2021). Managing the new digital workplace: Supercharging teams in the digital workplace. Deloitte Insights. Retrieved from https://www.deloitte.com/us/en/insights/topics/talent/supercharging-teams-in-the-digital-workplace.html

Edmondson, A. C. (1999). Psychological safety and learning behavior in work teams. Administrative Science Quarterly, 44(2), 350–383. https://doi.org/10.2307/2666999

Edmondson, A. C., & Lei, Z. (2014). Psychological safety: The history, renaissance, and future of an interpersonal construct. Annual Review of Organizational Psychology and Organizational Behavior, 1(1), 23–43. https://doi.org/10.1146/annurev-orgpsych-031413-091305

Fast, N. J., Gruenfeld, D. H., Sivanathan, N., & Galinsky, A. D. (2012). Illusory control: A generative force behind power's far-reaching effects. Psychological Science, 20(4), 502–508.

Frazier, M. L., Fainshmidt, S., Klinger, R. L., Pezeshkan, A., & Vracheva, V. (2017). Psychological safety: A meta‑analytic review and extension. Personnel Psychology, 70(1), 113–165. https://doi.org/10.1111/peps.12183

Gelfand, M. J., Lim, B. C., & Raver, J. L. (2007). Culture and accountability in organizations: Variations in forms of social control across cultures. Human Resource Management Review, 14(1), 135–156.

Hofstede, G. (2001). Culture's consequences: Comparing values, behaviors, institutions, and organizations across nations (2nd ed.). Sage.

House, R. J., Hanges, P. J., Javidan, M., Dorfman, P. W., & Gupta, V. (2004). Culture, leadership, and organizations: The GLOBE study of 62 societies. Sage.

Im, I., Hong, S., & Kang, M. S. (2011). An international comparison of technology adoption: Testing the UTAUT model. Information & Management, 48(1), 1–8. https://doi.org/10.1016/j.im.2010.09.001

Jiang, L., & Probst, T. M. (2016). A multilevel examination of affective job insecurity climate on safety outcomes. Journal of Occupational Health Psychology, 21(3), 366–377. https://doi.org/10.1037/a0040203

Katz, D., & Kahn, R. L. (1978). The social psychology of organizations (2nd ed.). Wiley.

Kemp, A. (2025, November 8). Manager support drives employee AI adoption. Gallup. Retrieved from https://www.gallup.com/workplace/694682/manager-support-drives-employee-adoption.aspx

Kim, B.-J., Kim, M.-J., & Lee, J. (2025). The dark side of artificial intelligence adoption: Linking AI adoption to employee depression via psychological safety and ethical leadership. Humanities and Social Sciences Communications, 12, Article 704.
https://doi.org/10.1057/s41599-025-05040-2

Kim, E. S., Sherman, D. K., & Taylor, S. E. (2019). Culture and social support provision: Who gives what and why. Personality and Social Psychology Bulletin, 45(8), 1120–1136.

Lewis, W., Agarwal, R., & Sambamurthy, V. (2003). Sources of influence on beliefs about information technology use: An empirical study of knowledge workers. MIS Quarterly, 27(4), 657–678. https://doi.org/10.2307/30036552

Louis, M. R. (1980). Surprise and sense making: What newcomers experience in entering unfamiliar organizational settings. Administrative Science Quarterly, 25(2), 226–251.

Magee, J. C., & Galinsky, A. D. (2008). Social hierarchy: The self‑reinforcing nature of power and status. Academy of Management Annals, 2(1), 351–398.
https://doi.org/10.1080/19416520802211628

Pearson, B. L. (2025, April 25). Microsoft's strategy for driving AI adoption in engineering teams. LinearB Blog. Retrieved from
https://linearb.io/blog/microsoft-strategy-for-driving-ai-adoption

Shen, L., Qiu, N., & Wang, Z. (2025). Psychological safety and trust as drivers of teachers' continued use of AI tools in classrooms. Scientific Reports, 15, 31426.
https://doi.org/10.1038/s41598-025-13789-4

Srite, M., & Karahanna, E. (2006). The role of espoused national cultural values in technology acceptance. MIS Quarterly, 30(3), 679–704. https://doi.org/10.2307/25148745

Straub, D., Keil, M., & Brenner, W. (1997). Testing the technology acceptance model across cultures: A three-country study. Information & Management, 33(1), 1–11.

Tarafdar, M., Tu, Q., Ragu-Nathan, B. S., & Ragu-Nathan, T. S. (2010). The impact of

    technostress on role stress and productivity. Journal of Management Information

    Systems, 27(1), 303–334.

Taras, V., Steel, P., & Kirkman, B. L. (2016). Does country equate with culture? Beyond

    geography in the search for cultural boundaries. Management International Review,

    56(4), 455–487.

Tesluk, P. E., & Jacobs, R. R. (1998). Toward an integrated model of work experience. Personnel

    Psychology, 51(2), 321–355.

Tyre, M. J., & Orlikowski, W. J. (1994). Windows of opportunity: Temporal patterns of

    technological adaptation in organizations. Organization Science, 5(1), 98–118.

    https://doi.org/10.1287/orsc.5.1.98

Venkatesh, V., Morris, M. G., Davis, G. B., & Davis, F. D. (2003). User acceptance of

    information technology: Toward a unified view. MIS Quarterly, 27(3), 425–478.

    https://doi.org/10.2307/30036540

Appendix

*Psychological Safety Scale Items*

1.  In my work team, it is easy to discuss difficult issues and problems.

2.  I won't receive criticism if I admit to an error or mistake.

3.  It is easy to ask a member of this team for help.

4.  I feel safe offering new ideas, even if they aren't fully-formed plans.

5.  In my work team, people are accepted for being different.

6.  My teammates welcome my ideas and give them time and attention.

7.  Members of my work team could easily describe the value of others' contributions.

8.  In my work team, people talk about mistakes and ways to improve and learn from them.

9.  We take time to find new ways to improve our team's work processes.

10. Members of my work team raise concerns they have about team plans or decisions.

11. We try to discover our underlying assumptions and seek counterarguments about issues under discussion.